\documentclass[11pt]{article}
\usepackage[totalwidth=450pt,totalheight=584pt]{geometry}
\usepackage{amsmath,amssymb,epsf,graphics,graphicx}
\usepackage{psfrag}
\usepackage{amsfonts}
\usepackage{mathrsfs}
\usepackage{epsfig}
\usepackage{relsize}
\usepackage{epstopdf}

\def\sl(2){\alg{sl}(2)}

\def\a {\alpha}

\def\la{\label}

\def\ov{\over}

\def\tp{{\widetilde p}}

\newcommand{\alg}[1]{\mathfrak{#1}}

\newcommand{\atopfrac}[2]{\genfrac{}{}{0pt}{}{#1}{#2}}

\newcommand{\bem}{\left (\begin{matrix}}
\newcommand{\eem}{\end{matrix} \right )}

\newcommand{\nn}{\nonumber}

\newcommand{\be}{\begin{equation}}
\newcommand{\ee}{\end{equation}}
\newcommand{\ba}{\begin{eqnarray}}
\newcommand{\ea}{\end{eqnarray}}

\newcommand\blank[1]{}

\newcommand{\fract}[2]{{\textstyle\frac{#1}{#2}}}

\newcommand\eq{\begin{equation}}
\newcommand\en{\end{equation}}
\newcommand\bea{\begin{eqnarray}}
\newcommand\eea{\end{eqnarray}}
\newcommand{\resection}[1]{\setcounter{equation}{0}\section{#1}}

\newcommand\ep{\epsilon}

\begin{document}
\begin{titlepage}
\vskip 1.2cm
\begin{center}
{\Large{\bf Thermodynamic Bethe Ansatz for planar AdS/CFT: a proposal}}
\end{center}
\vskip 0.8cm
\centerline{Diego Bombardelli$^1$,
Davide Fioravanti$^1$ and Roberto Tateo$^2$}
\vskip 0.9cm

\centerline{${}^1$ \small INFN-Bologna and Dipartimento di Fisica,
Universit\`a di Bologna,
}
\centerline{\small Via Irnerio 46, Bologna, Italy}
\vskip 0.2cm
\centerline{${}^{2}$ \small Dip.\ di Fisica Teorica and INFN,
Universit\`a di Torino,}
\centerline{\small Via P.\ Giuria 1, 10125 Torino, Italy}
\vskip 0.2cm
\centerline{e-mails:}
\centerline{bombardelli@bo.infn.it, fioravanti@bo.infn.it, tateo@to.infn.it}

\vskip 1.25cm
\begin{abstract}
\noindent
Moving from the mirror theory Bethe-Yang equations proposed by Arutyunov and Frolov, we derive the thermodynamic Bethe Ansatz equations which should control the spectrum of the planar  $\text{AdS}_5/\text{CFT}_4$ correspondence. The associated set of universal functional relations (Y-system) satisfied by the exponentials of the TBA pseudoenergies  is deduced, confirming the structure inferred by  Gromov, Kazakov and Vieira.
\end{abstract}
\end{titlepage}
\setcounter{footnote}{0}
%
\resection{A bird's-eye view between integrability and AdS/CFT}

A very peculiar phenomenon in modern theoretical physics has been taking place at the encounter of two branches: on one side the subject of quantum/statistical two-dimensional integrability \cite{INT} and on the other the gauge/string correspondences \cite{M-GKP-W} in their planar case. Actually, the entrance of  integrability into the realm of reggeised gluons of infinite colour QCD in its leading logarithmic approximation was already observed by Lipatov in~\cite{L}.

More specifically, the AdS/CFT conjecture relates, by a strong/weak coupling duality, a type IIB superstring theory on the curved space-time $\text{AdS}_5\times\text{S}^5$ and the
conformal ${\cal N}=4$ Super Yang-Mills theory (SYM) in four dimensions  on the boundary of $\text {AdS}_5$ \cite{M-GKP-W}.  As a consequence and particular case, the energy of a specific string state ought to be equal the anomalous dimension of the corresponding local gauge invariant operator in the
quantum field theory. Yet, the mechanism of integrability in this triadic relation is not fully understood. For sure, the discovery of integrability in the classical string theory was a great achievement  \cite{BPR}, both from the conceptual and the practical (i.e. calculative) point of view.

At the other side of the correspondence, in the maximally SYM theory for colour number $N\rightarrow \infty$ so that the 't Hooft coupling $Ng_{YM}^2=\lambda = 4 \pi^2 g^2$, with $g$ proportional to free string tension, is kept fixed, only the planar Feynman diagrams and single trace composite operators survive. Besides the pioneering interpretation of \cite{L2} in terms of a $sl(2)$ spin chain (in the QCD case), the constituent operators in the purely scalar sector at one loop have been unveil to correspond to the degrees of freedom of an integrable $so(6)$ spin chain, thus making the mixing matrix (or dilatation operator) to coincide with this integrable $so(6)$ spin Hamiltonian \cite{MZ}. Being integrable, the spectrum of this Hamiltonian comes out by means of the Bethe Ansatz (BA) (in one of its various forms) \cite{INT} and described by the so-called Bethe Ansatz equations for the 'rapidities' which parametrise
the operators in the trace. Albeit a description of the dilatation operator at all loops as a spin chain Hamiltonian is still missing, the integrability has been showing up in the form of spin-chain-like Bethe equations (for $g$ dependent rapidities still parametrising the operators in the trace, likewise to the one-loop case), which are valid at least in the {\it asymptotic regime} of large quantum numbers (cf. below) . Eventually, a set of equations for the whole theory has been proposed by Beisert and Staudacher \cite{BS}. Computationally, the BA energy, $E(g)$, yields the anomalous part of the conformal dimension
\begin{equation}
\Delta = \Delta_{bare}+g^2 E(g) \, , \label{Delta}
\end{equation}
where $\Delta_{bare}$ is the bare or classical dimension. As said before, this quantity must also be given by the quantum energy of a suitable string state ($E_{string}=\Delta$).  By a semiclassical procedure on the string sigma model, this fact has opened a road to fix a phase factor, the so-called dressing factor, entering the Bethe equations (and the $S$-matrix) \cite{AFS,HL,BHL,BES}. Of course, $\Delta$, $\Delta_{bare}$ and $E(g)$ may depend also on other quantum numbers, like the spin chain length $L$, -- which also plays the r\^ole of a string angular momentum --, other angular momenta, the Lorentz spin s, etc.. Yet, the Beisert-Staudacher equations enjoy a validity seriously restricted by their scattering matrix origin, namely the length $L$ and other quantum numbers need to be large. More precisely,  starting from a certain loop order these equations are plagued by the so-called `wrapping problem'~\cite{Sieg:2005kd, AJK}. Nevertheless, as scattering $S$-matrix equati!
 ons \cite{Staudacher:2004tk}, they are indeed correct and they can be interpreted as Bethe-Yang quantisation conditions~\cite{Be-MM} \cite{Be-H}.

In quantum integrable 2D relativistic massive field theories the problem of deriving off-shell quantities from on-shell information has been already addressed in many cases. For the purpose of this paper it is relevant  the derivation by Al. B. Zamolodchikov of the finite-size  ground state energy  from  the $S$-matrix \cite{Zamolodchikov:1989cf}. Let us define the  theory on a torus space-time geometry. The space direction is finite with circumference $L$, time is periodic with period $R\rightarrow\infty$. Zamolodchikov's fascinating idea is to exchange space and time by defining a {\it mirror theory} in the infinite space $R$. In this mirror theory  the space interval is infinite and the asymptotic Bethe-Yang equations hold true, but time is compact with size $L$. Now, we may interpret $L=1/T$ as the inverse temperature and use the Yang-Yang thermodynamic Bethe Ansatz (TBA) procedure \cite{YY} to find the minimum free energy or equivalently the ground state energy for the !
 (original) {\it direct} theory on a space circumference  with size $L$. In the following, we will extend  this procedure to the non-relativistic case relevant for  the AdS/CFT correspondence.

We have been convinced that this strategy may be successful also in a complicated non-relativistic theory such as the AdS/CFT correspondence by the recent striking confirmation due to a sort of ancestor of the TBA for relativistic quantum field theory. In fact, L\"{u}scher developed a method  to compute, from scattering data, the finite-size corrections to the mass gap~\cite{Luscher}. Later on, this method was specifically applied to integrable quantum field theories~\cite{KM} and revealed itself as the leading term in the TBA large size expansion~\cite{Bazhanov:1996aq, Dorey:1996re}. Recently, a sophisticated extension of these ideas to the AdS/CFT correspondence has given  striking results for the Konishi operator at four loops \cite{BJ} and an impressive confirmation of the perturbative computations of \cite{FSSZ}.

In this article, we will start from the equations recently formulated by Arutyunov and Frolov in \cite{Arutyunov:2009zu} for the mirror theory of the $\text{AdS}_5\times\text{S}^5$ superstring theory. These equations are derived by implementing the classification of all the particles and bound states in the Bethe-Yang equations derived in \cite{Arutyunov:2007tc}. The classification is obtained with the formulation of the so-called string hypothesis of the Hubbard model (cf. \cite{Takahashi}): the map of the direct theory equations \cite{Be-MM} into those of Hubbard's was already observed by Beisert \cite{Be-H}. Initially, we will modify the equations -- in analogy with those of the Hubbard model \cite{Takahashi} --,  so that we can take into account the information on the so-called $k-\Lambda$ strings. In this way, we produce a complete set of string equations for  implementing the
thermodynamic Bethe Ansatz method and  derive a set of TBA  equations for the
single particle dressed energies (the pseudoenergies). As a conclusion, the pseudoenergies determine the (free) energy via a non-linear integral functional. We shall make explicit the similarity between  our TBA equations and  those for the Hubbard model and then derive a universal system of functional relations (the Y-system) for the exponential of the pseudoenergies. The universality of a Y-system consists in the fact that, at least for relativistic theories, it is the same for the excited states as well. Yet, there is by now a consolidated way towards excited states in relativistic massive field theories \cite{Bazhanov:1996aq, Dorey:1996re, FMQR}. A very brief description of this procedure for the present case will be sketched in the final section, with the aim to gain a better control of the energy/dimension spectrum of the $\text{AdS}_5/\text{CFT}_4$ correspondence for any value of the coupling constant $g$ and even for {\it short} operators. Apparently, the Y-system st!
 ructure matches that recently proposed by Gromov, Kazakov and Vieira \cite{Gromov:2009tv}.

\resection{The equations for the root densities}

As anticipated before, we need to pass from the $\text{AdS}_5\times \text{S}^5$ theory defined on a circumference of length $L$ to its mirror and this has been extensively investigated by Arutyunov and Frolov since the paper~\cite{Arutyunov:2007tc}. In particular, they derive from the $S$-matrix the Bethe-Yang equations for the fundamental particles of the mirror theory. Then, more recently \cite{Arutyunov:2009zu}, they extend these equations also to $Q$-particle bound states of the $\text{AdS}_5\times \text{S}^5$ mirror theory in the form
\begin{eqnarray}\la{BY}
e^{i\tp_{k} R} &=&\prod_ {\textstyle\atopfrac{l=1}{l\neq k}}
^{K^{\mathrm{I}}}\big(S_0(\tp_{k},\tp_{l})\big)^2
\prod_{\a=1}^{2}\prod_{l=1}^{K^{\mathrm{II}}_{(\a)}}\frac{{x_{k}^{+}-y_{l}^{(\a)}}}{x_{k}^{-}-y_{l}^{(\a)}}
\sqrt{\frac{x_k^-}{x_k^+}} \nonumber~,\\
-1&=&\prod_{l=1}^{K^{\mathrm{I}}}\frac{y_{k}^{(\a)}-x^{+}_{l}}{y_{k}^{(\a)}-x^{-}_{l}}\sqrt{\frac{x_l^-}{x_l^+}}
\prod_{l=1}^{K^{\mathrm{III}}_{(\a)}}\frac{v_{k}^{(\a)}-w_{l}^{(\a)}+\frac{i}{g}}{v_{k}^{(\a)}-w_{l}^{(\a)}-\frac{i}{g}} ~,\\
\nonumber
1&=&\prod_{l=1}^{K^{\mathrm{II}}_{(\a)}}\frac{w_{k}^{(\a)}-v_{l}^{(\a)}-\frac{i}{g}}{w_{k}^{(\a)}-v_{l}^{(\a)}+\frac{i}{g}}
\prod_ {\textstyle\atopfrac{l=1}{l\neq
k}}^{K^{\mathrm{III}}_{(\a)}}\frac{w_{k}^{(\a)}-w_{l}^{(\a)}+\frac{2i}{g}}{w_{k}^{(\a)}-w_{l}^{(\a)}-\frac{2i}{g}}~,
\end{eqnarray}
where

\be
\big(S_0(\tp_{k},\tp_{l})\big)^2=\frac{x_k^--x_l^+}{x_k^+-x_l^-}\frac{1-\frac{1}{x_k^+x_l^-}}{1-\frac{1}{x_k^-x_l^+}}\,\sigma^2(x_k,x_l)
\ee
is the $a=0$ light-cone gauge scalar factor of the mirror $S$matrix, with $\sigma(x_k,x_l)$ the dressing factor in the mirror theory \cite{Arutyunov:2007tc}.
Thanks to a so-far formal resemblance of the last two BA Equations (BAEs) with those of a inhomogeneous Hubbard model, they can formulate a string hypothesis for the solutions, in strict analogy with the Takahashi' s one~\cite{Takahashi}. In few words, we assume that the thermodynamically relevant solutions \footnote{There is no definitive proof of the string hypothesis, though it seems to give always the correct thermodynamic limit. There might well be other kinds of solutions (which should not affect the thermodynamics).}  of (\ref{BY}) in the limit of large $R, K^I, K^{II}_{(\alpha)}, K^{III}_{(\alpha)}$ rearrange themselves into complexes -- the so-called strings -- with real centers and all the other complex roots symmetrically distributed around these centers along the imaginary direction. Paying attention to the presence of two coupled Hubbard models for $\alpha=1,2$, the strings may be classified as follows:
\ba
\label{strings}
&&\mbox{1) $N_Q$ $Q$-particles with real momenta $\tilde{p}^Q_k$ and real rapidities $u_k^Q$}:\nonumber\\
&&\ \ \ \ u_k^{Q,j}=u_k^Q+(Q+1-2j)\frac{i}{g}\,,\ \ j=1,...,Q\ ;\\
&&\mbox{2) $N_y^{(\a)}$ $y^{(\a)}$-particles with real momenta $q_k^{(\a)}$;}\nonumber\\
&&\mbox{3) $N_{M|v}^{(\a)}$ $vw$-strings with real centers $v_k^M$, $2M$ roots of type $v$ and M of type $w$:}\ \ \nonumber\\
&&\ \ \ \ v_k^{M,j}=v_k^M\pm(M+2-2j)\frac{i}{g}\,,\ \ j=1,...,M\ ;\\
&&\ \ \ \ w_k^{M,j}=v_k^M+(M+1-2j)\frac{i}{g}\,,\ \ j=1,...,M\ ;\\
&&\mbox{4) $N_{N|w}^{(\a)}$ $w$-strings with real centers $w_k^N$ and $N$ roots of type $w$:}\nonumber\\
&&\ \ \ \ w_k^{N,j}=w_k^N+(N+1-2j)\frac{i}{g}\,,\ \ j=1,...,N\ .
\ea
If the variables $u_k, v_k$ and $w_k$ in (\ref{BY}) are replaced by $u_k^{Q,j}, v_k^{M,j}, w_k^{M,j}$ and $w_k^{N,j}$, and the products on the internal string index $j$ are made, then the equations for the real centers of the various kinds of string (\ref{strings}) can be recast into the  following form~ \cite{Arutyunov:2009zu}

\ba
&&1=e^{i\tilde{p}_{k}^Q R}\prod_{Q'=1}^{\infty}\prod_{\textstyle\atopfrac{l=1}{l\neq k}}
^{N_{Q'}}S_{\sl(2)}^{QQ'}(x_{k},x_{l})
\prod_{\a=1}^{2}\prod_{l=1}^{N_y^{(\a)}}\frac{{x_{k}^--y_{l}^{(\a)}}}{x_{k}^+-y_{l}^{(\a)}}
\sqrt{\frac{x_k^+}{x_k^-}}\, \prod_{M=1}^{\infty} \prod_{l=1}^{N_{M|vw}^{(\a)}}S_{xv}^{QM}(x_{k},v_{l,M}^{(\a)})\,,\label{string1}\\
&&-1=\prod_{Q=1}^{\infty}\prod_{l=1}^{N_Q}\frac{y_{k}^{(\a)}-x^+_{l}}{y_{k}^{(\a)}-x^-_{l}}\sqrt{\frac{x_l^-}{x_l^+}}
\prod_{M=1}^{\infty} \prod_{l=1}^{N_{M|vw}^{(\a)}}\frac{v_{k}^{(\a)}-v_{l,M}^{(\a)-}}{v_{k}^{(\a)}-v_{l,M}^{(\a)+}} \,\prod_{N=1}^{\infty} \prod_{l=1}^{N_{N|w}^{(\a)}}\frac{v_{k}^{(\a)}-w_{l,N}^{(\a)-}}{v_{k}^{(\a)}-w_{l,N}^{(\a)+}} \,,\label{string2}\\
&&\prod_{Q=1}^{\infty}\prod_{l=1}^{N_Q}S_{xv}^{QK}(x_{l},v_{k,K}^{(\a)})=\prod_{M=1}^{\infty} \prod_{l=1}^{N_{M|vw}^{(\a)}}S_{vv}^{KM}(v_{k,K}^{(\a)},v_{l,M}^{(\a)})\prod_ {N=1}^{\infty} \prod_{l=1}^{N_{N|w}^{(\a)}}S_{vw}^{KN}(v_{k,K}^{(\a)},w_{l,N}^{(\a)})\,,\label{string3}\\
&&(-1)^K=\prod_{l=1}^{N_y^{(\a)}}\frac{w_{k,K}^{(\a)-}-v_{l}^{(\a)}}{w_{k,K}^{(\a)+}-v_{l}^{(\a)}}
\prod_ {N=1}^{\infty} \prod_{l=1}^{N_{N|w}^{(\a)}}S_{ww}^{KN}(w_{k,K}^{(\a)},w_{l,N}^{(\a)}) \,,~~~~~\label{string4}
\ea
where, for shortness' sake, all the $x$-variables have to be read as
\be
x_k^{\pm}\equiv x_k^{Q\pm}=x\left(u_k^Q \pm i\frac{Q}{g}\right)\,,
\ee
and the definitions of the variables $x^{\pm}, v, v_K^{\pm}$ and $w_K|{\pm}$ are reported in Appendix A.
The $S$-matrices are defined as follows:

\ba
&&S_{\sl(2)}^{QQ'}(x_{k},x_{l})=\left(\frac{x_k^{Q+}-x_l^{Q'-}}{x_k^{Q-}-x_l^{Q'+}} \right) \left(\frac{1-\frac{1}{x_k^{Q-}x_l^{Q'+}}}{1-\frac{1}{x_k^{Q+}x_l^{Q'-}}} \right)\sigma(x^{Q\pm}_k,x^{Q'\pm}_l)^{-2}~,\\
&&S_{xv}^{QM}(x_k,v_{l,M})=\left(\frac{x_k^{Q-}-x(v_{l,M}^+)}{x_k^{Q+}-x(v_{l,M}^+)} \right) \left(\frac{x_k^{Q-}-x(v_{l,M}^-)}{x_k^{Q+}-x(v_{l,M}^-)}\right) \left(\frac{x_k^{Q+}}{x_k^{Q-}} \right)\prod_{j=1}^{M-1} \left(\frac{u_k^Q-v_{l,M}-i\frac{Q-M+2j}{g}}{u_k^Q-v_{l,M}+i\frac{Q-M+2j}{g}} \right)~,\nonumber\\
&&S_{vv}^{KM}(x,y)=S_{vw}^{KM}(x,y)=S_{ww}^{KM}(x,y)= S_{KM}(x-y)~,\nonumber\\
&&S_{KM}(u)= \left( {u + i\frac{|K-M|}{ g} \over u -i\frac{|K-M|}{g}} \right)
\left( {u + i\frac{K+M}{g} \over u -i \frac{K+M}{g}} \right)
\prod_{k=1}^{\text{min}(K,M)-1} \left( {u + i\frac{|K-M|+2k}{g} \over u -i\frac{|K-M|+2k}{g}} \right)^2~,
\label{SKM}
\ea
where $S_{\sl(2)}^{QQ'}(x_{k},x_{l})$ is obtained from $\big(S_0(\tp_{k},\tp_{l})\big)^2$ and the fusion procedure \cite{Roiban:2006gs, Chen:2006gq}.
Now, a simple crucial observation enters the stage: the last term in the r.h.s. of (\ref{string3}) fails the resemblance with the usual Hubbard BAEs implemented by string hypothesis  \cite{Takahashi, onedHubbard}. In fact, we need one more step: we can easily see that the equation for the $vw$ strings -- corresponding to the Hubbard $k-\Lambda$ strings -- do not have in the r.h.s. a term of interaction between the $w$ and $vw$ strings; on the contrary there is a scattering term between a $vw$ string and a single $v^{(\alpha)}$ (which do not belong to any string, but its own). Therefore, we may derive an intermediate equation
\begin{eqnarray}\la{BEw3}
-1&=&\prod_{l=1}^{N_y^{(\a)}}\frac{w_{k}^{(\a)}-v_{l}^{(\a)}-\frac{i}{g}}{w_{k}^{(\a)}-v_{l}^{(\a)}+\frac{i}{g}}
\prod_ {N=1}^{\infty} \prod_{l=1}^{N_{N|w}^{(\a)}}\frac{w_{k}^{(\a)}-w_{l,N}^{(\a)-} +\frac{i}{g}} {w_{k}^{(\a)}-w_{l,N}^{(\a)+}+\frac{i}{g}} \,\frac{w_{k}^{(\a)}-w_{l,N}^{(\a)-} -\frac{i}{g}}  {w_{k}^{(\a)}-w_{l,N}^{(\a)+}-\frac{i}{g}}\,,~~~~~
\end{eqnarray}
and choose $w_{k}^{(\a)}$ belonging to a $vw$-string. With this little trick \footnote{After the first version of this paper appeared on the arXiv, this trick was implemented in a revised version of \cite{Arutyunov:2009zu}}, we  obtain
\be
\prod_{N=1}^{\infty} \prod_{l=1}^{N_{N|w}^{(\a)}}S_{vw}^{KN}(v_{k,K}^{(\a)},w_{l,N}^{(\a)})=(-1)^K\prod_{l=1}^{N_y^{(\a)}}\frac{v_{k.K}^{(\a)+}-v_l^{(\a)}}{v_{k.K}^{(\a)-}-v_l^{(\a)}}
~,
\ee
and  finally we can rewrite (\ref{string3}) in a form re-echoing the Hubbard one
\be
\prod_{Q=1}^{\infty}\prod_{l=1}^{N_Q}S_{xv}^{QK}(x_{l},v_{k,K}^{(\a)})=\prod_{M=1}^{\infty} \prod_{l=1}^{N_{M|vw}^{(\a)}}S_{vv}^{KM}(v_{k,K}^{(\a)},v_{l,M}^{(\a)})\prod_ {N=1}^{\infty} \prod_{l=1}^{N_{y}^{(\a)}}S_{vy}^{K}(v_{k,K}^{(\a)},v_{l}^{(\a)})\,.
\label{newSM}
\ee
In (\ref{newSM})  we have introduced  a new scattering matrix
\be
S_{vy}^{K}(v_{k,K}^{(\a)},v_{l}^{(\a)})=\frac{v_{k.K}^{(\a)+}-v_l^{(\a)}}{v_{k.K}^{(\a)-}-v_l^{(\a)}}=\frac{v_{k.K}^{(\a)}-v_l^{(\a)}+iK/g}{v_{k.K}^{(\a)}-v_l^{(\a)}-iK/g}\ .
\ee

At this point, we can follow  the standard TBA procedure \cite{YY, Takahashi, onedHubbard, Zamolodchikov:1989cf}, which goes in a very sketchy way as follows. After taking the logarithm of these equations, we shall consider the thermodynamic limit ($K^I, N_y^{(\alpha)}, N_{vw}^{(\alpha)}, N_{w}^{(\alpha)}, R\rightarrow\infty$) while keeping  the densities finite (sums of root and hole densities, respectively)
\ba
\rho_Q(\tilde{p})&=&\rho^r_Q(\tilde{p})+\rho^h_Q(\tilde{p})=\lim_{R\rightarrow\infty}\frac{I^Q_{k+1}-I^Q_k}{R (\tilde{p}^Q_{k+1}-\tilde{p}^Q_k)}\ ,\\
\rho^{\a}_y(q)&=&\rho^{r\a}_y(q)+\rho^{h\a}_y(q)=\lim_{R\rightarrow\infty}\frac{I'^{\a}_{k+1}-I'^{\a}_k}{R (q^{(\a)}_{k+1}-q^{(\a)}_k)}\ ,\\
\rho^{\a}_{v,K}(\lambda)&=&\rho^{r\a}_{v,K}(\lambda)+\rho^{h\a}_{v,K}(\lambda)=\lim_{R\rightarrow\infty}\frac{J^{K^{\a}}_{k+1}-J^{K^{\a}}_k}{R (\lambda^{(\a)}_{k+1}-\lambda^{(\a)}_k)}\ ,\\
\rho^{\a}_{w,K}(\lambda)&=&\rho^{r\a}_{w,K}(\lambda)+\rho^{h\a}_{w,K}(\lambda)=\lim_{R\rightarrow\infty}\frac{J'^{K^{\a}}_{k+1}-J'^{K^{\a}}_k}{R (\lambda^{(\a)}_{k+1}-\lambda^{(\a)}_k)}\  ,
\ea
where the $I$s and the $J$s are the integer and half-integer quantum numbers. Eventually, we can produce for the thermodynamic state the following integral equations constraining the densities
\ba
\label{dens}
\rho_Q(\tilde{p})&=&\frac{1}{2\pi}+\sum_{Q'=1}^{\infty}(\phi_{\sl(2)}^{QQ'}*\rho_{Q'}^r)(\tilde{p})+\sum_{\alpha=1}^2\left[(\phi_{xy}^{Q}*\rho_y^{r\a})+\sum_{M=1}^{\infty}(\phi_{xv}^{QM}*\rho_{v,M}^{r\a})\right](\tilde{p})\ ,~~~~~~\\
\rho_y^{\a}(q)&=&\sum_{Q=1}^{\infty}(\phi_{yx}^{Q}*\rho_Q^{r})(q)+\sum_{M=1}^{\infty}(\phi_{yv}^{M}*\rho_{v,M}^{r\a})(q)+\sum_{N=1}^{\infty}(\phi_{yw}^{N}*\rho_{w,N}^{r\a})(q)\ ,\\
\rho^{\a}_{v,K}(\lambda)&=&\sum_{M=1}^{\infty}(\phi_{vv}^{KM}*\rho_{v,M}^{r\a})(\lambda)+(\phi_{vx}^{KQ}*\rho_Q^{r})(\lambda)+(\phi_{vy}^{K}*\rho_{y}^{r\a})(\lambda)\ ,\\
\rho^{\a}_{w,K}(\lambda)&=&\sum_{M=1}^{\infty}(\phi_{ww}^{KM}*\rho_{w,M}^{r\a})(\lambda)+(\phi_{wy}^{K}*\rho_y^{r\a})(\lambda)\ ,
\label{dens2}
\ea
where the symbol * denotes the {\it usual} convolution (on the second variable) $(\phi*g)(z)=\int dz'\,\phi(z,z')\,g(z')$ and the kernels are defined in Appendix A \footnote{We begin to notice that here the kernels $\phi(z,z')$ do not necessarily depend on the difference $(z-z')$.}.

\resection{The thermodynamic Bethe Ansatz equations}
\label{TBAR}

We continue our very sketchy presentation of the derivation of the TBA equations. For this purpose,
we  express the entropy in terms of the hole and root densities \footnote{Hereafter the integration measure $d\tilde{p}$ has to be interpreted as Stieltjes measure $\frac{d\tilde{p}}{du}\,du$, as $\tilde{p}$ depends on (the parameters) $Q$ and $g$ as well.} ($\rho^h$ and $\rho^r$, respectively)
\ba
S&=&\sum_{Q=1}^{\infty}\int_{-\infty}^{\infty}d\tilde{p} \left([\rho^r_Q(\tilde{p})+\rho^h_Q(\tilde{p})]\ln[\rho^r_Q(\tilde{p}) +\rho^h_Q(\tilde{p})]
-\rho^r_Q(\tilde{p})\ln\rho^r_Q(\tilde{p})-\rho^h_Q(\tilde{p})\ln\rho^h_Q(\tilde{p}) \right) \nn\\
 &+&\sum_{\a=1}^{2}\int_{-\pi}^{\pi}dq \left([\rho^{r\a}_y(q)+\rho^{h\a}_y(q)]\ln[\rho^{r\a}_y(q)
 +\rho^{h\a}_y(q)]-\rho^{r\a}_y(q)\ln\rho^{r\a}_y(q)-\rho^{h\a}_y(q)\ln\rho^{h\a}_y(q) \right) \nn\\
&+&\sum_{\a=1}^{2}\sum_{M=1}^{\infty}\int_{-\infty}^{\infty}d\lambda \Big( [\rho^{r\a}_{v,M}(\lambda)+\rho^{h\a}_{v,M}(\lambda)]\ln[\rho^{r\a}_{v,M}(\lambda)
+\rho^{h\a}_{v,M}(\lambda)]-\rho^{r\a}_{v,M}(\lambda)\ln\rho^{r\a}_{v,M}(\lambda) \nonumber\\
&-&\rho^{h\a}_{v,M}(\lambda)\ln\rho^{h\a}_{v,M}(\lambda) \Big) \nn \\
&+&\sum_{\a=1}^{2}\sum_{N=1}^{\infty}\int_{-\infty}^{\infty}d\lambda \Big([\rho^{r\a}_{w,N}(\lambda)+\rho^{h\a}_{w,N}(\lambda)]\ln[\rho^{r\a}_{w,N}(\lambda)
+\rho^{h\a}_{w,N}(\lambda)]-\rho^{r\a}_{w,N}(\lambda)\ln\rho^{r\a}_{w,N}(\lambda)\nonumber\\
&-&\rho^{h\a}_{w,N}(\lambda)\ln\rho^{h\a}_{w,N}(\lambda) \Big)~,
\ea
and then minimise the free energy per unit length
\be
f(T)=\tilde{H}-TS~,
\label{rhofree}
\ee
where $\tilde{H}$ is the mirror energy per unit length~\cite{Arutyunov:2007tc}:
\be
\tilde{H}=2 \sum_{Q=1}^{\infty}\int_{-\infty}^{\infty}d\tilde{p}~\mbox{arcsinh}\left(\fract{\sqrt{Q^2+\tilde{p}^2}}{2g}\right)\rho_Q^r(\tilde{p})\ .
\ee
As stated before, then we ought to take as temperature $T$ of the mirror theory the inverse of the size $L$ in the AdS/CFT: $T=1/L$. The extremum condition $\delta f=0$ under the constraints (\ref{dens})-(\ref{dens2}) entails the final set of thermodynamic Bethe Ansatz equations for the pseudoenergies $\ep_A$ such that
\eq
\epsilon_A=\ln {\rho^h_A \over \rho^r_A}~,~~~{1 \over e^{\ep_A}+1}= {\rho^r_A \over \rho_A}~,~~~
L_A=\ln\left(1+e^{-\ep_A} \right)~,
\label{dfde}
\en
with the short indication of the collective index $A$ for the different density labels.
The ground state thermodynamic Bethe Ansatz  equations are
\ba
\epsilon_Q(\tilde{p})&=&2L\,\mbox{arcsinh}
\left(\frac{\sqrt{Q^2-\tilde{p}^2}}{2g}\right)
-\sum_{Q'=1}^{\infty}(\phi_{\sl(2)}^{Q'Q}*L_{Q'})(\tilde{p})\nn\\
&-&\sum_{\a=1}^2(\phi_{yx}^{Q}*L^{\a}_y)(\tilde{p})
-\sum_{\a=1}^2\sum_{M=1}^{\infty}(\phi_{vx}^{MQ}*L^{\a}_{v,M})(\tilde{p})~,~~~\label{TBA1} \\
\epsilon_y^{\alpha}(q)&=&-\sum_{Q=1}^{\infty}(\phi_{xy}^{Q}*L_{Q})(q)
- \sum_{M=1}^{\infty} (\phi_{wy}^{M}*L^{\alpha}_{w,M})(q) \nn \\
&-&\sum_{N=1}^{\infty}(\phi_{vy}^{N}*L^{\alpha}_{v,N})(q)~,~~~ \label{TBA2}\\
\epsilon_{v,K}^{\alpha}(\lambda)&=&-
\sum_{Q=1}^{\infty}(\phi_{xv}^{QK}*L_{Q})(\lambda)-(\phi_{yv}^{K}*L_{y}^{\alpha})(\lambda)  \nn \\
&-&\sum_{M=1}^{\infty}(\phi_{vv}^{MK}*L_{v,M}^{\alpha})(\lambda)~,~~~\label{TBA3} \\
\epsilon^{\alpha}_{w,K}(\lambda)&=&-(\phi_{yw}^{K}*L^{\alpha}_y)(\lambda)
-\sum_{M=1}^{\infty}(\phi_{ww}^{MK}*L^{\alpha}_{w,M})(\lambda)~,~~~\label{TBA4}
\ea
with $\alpha=1,2$~,  $Q=1,2,\dots$ and $K=1,2,\dots$ . Notice that, apart from the specific form of the kernels (see Appendix~A for their definitions), the TBA equations are similar in form to
the density equations (\ref{dens}-\ref{dens2}), provided we exchange $\rho\rightarrow -L$. However, we should stress that  on our way from (\ref{dens}-\ref{dens2}) to (\ref{TBA1}-\ref{TBA4}) we have made an abuse of notation and changed definition for the convolution $*$ moving on to the first variable
\eq
(\phi*g)(z)= \int dz' \, \phi(z',z)\,g(z')~.
\en
When the kernel $\phi(z,z')$ depends only on the difference $|z-z'|$, as for example in the
relativistic theories discussed in ~\cite{Zamolodchikov:1989cf, Klassen:1990dx}, this change in the definition of  * can be avoided by keeping the convolution on the second variable. However, in the present framework some of the kernels have a genuinely different functional dependence on the two independent variables and this simplification is absent. Moreover, an important comment is here on the integration limits: they are
from  $-\infty$ to $\infty$ for the $\lambda$- and $\tilde{p}$-variables but from $-\pi$  to $\pi$ for the $q$-variables.

As a concluding result, the {\it minimal} free energy for the mirror theory results by inserting the TBA equations  into the general (\ref{rhofree}) and is given by the following non-linear functional of the pseudoenergies $\epsilon_Q(u)$
\ba
f(T)=-T\sum_{Q=1}^{\infty}\int_{-\infty}^{\infty}\frac{d\tilde{p}}{2\pi}\ln(1+e^{-\epsilon_Q(\tilde{p})})
=-T\sum_{Q=1}^{\infty}\int_{\infty}^{\infty}\frac{du}{2\pi}\frac{d\tilde{p}}{du}\ln(1+e^{-\epsilon_Q(u)})~.
\ea
Consequently, the ground state energy for the AdS/CFT theory on a circumference with length $L=1/T$ ought to satisfy the relation
\eq
E_0(L)=Lf(1/L)~.
\en

As we have kept the total densities finite, it is natural to introduce chemical potentials $\mu_A$. This has been already finalised in relativistic theories by ~\cite{Klassen:1990dx}.
The TBA equations (\ref{TBA1}--\ref{TBA4}) do not change their form, but for this simple replacement
\eq
L_A=\ln (1+ e^{-\epsilon_A})  \rightarrow L_{A,\lambda}=\ln (1+ \lambda_A e^{-\epsilon_A})~,
\label{Ldef}
\en
involving the fugacities $\lambda_A=e^{\mu_A/T}$. Here, we would like to conjecture that their introduction should be related to the zero energy  of the ground state (independently of the value of $T$) which is a half BPS protected state. It is a consequence of a result by ~\cite{Martins:1991hw}, further developed  in~\cite{Fendley:1991xn} and in \cite{Fendley:1991ve} that in particular ${\cal N}=2$ supersymmetric theories this size invariant state can be selected via a suitable tuning of the  TBA fugacities. A plot describing this interesting transition, as the fugacities  approach  these critical values can be found in \cite{Dorey:2002gj}.
In our case we expect zero energy as soon as the fugacities reach these values
\eq
\lambda_Q=1~,~~\lambda_{v,K}^{\alpha}=-1~,~~\lambda_{w,K}^{\alpha}=(-1)^{K+1}~,
~~~\lambda_y^{\alpha}=-1~,~
\label{fugacities}
\en
$(\alpha=1,2~,K=1,2,\dots)$.

Physically, this modification corresponds to the calculation of the Witten index. In (\ref{fugacities}), the fermionic and bosonic character of the pseudoparticles is chosen following an analogy with other scattering-matrix  models and considering the evident $Z_2$-symmetry of the TBA equations.
There are -of course- other possibilities. The vanishing of
ground state energy in TBA models is a very  delicate issue and we  prefer to postpone this discussion to the near future and in presence  of    analytic or numerical evidences.

\subsection{A comparison with the Hubbard TBA equations}

As the reader can see in  Appendix~A, some kernels in (\ref{TBA1})-(\ref{TBA4}) actually depends on the difference of rapidities. Therefore, the convolutions involving these kernels is a  standard  `difference' convolution, i.e. $(f*g)(z)=\int dz'\,f(z-z')\,g(z')$. In other words, we may rewrite the equations (\ref{TBA2})-(\ref{TBA4}) in a form that is closer to the TBA equations of the Hubbard model, as we might expect from the analogy at the level of Bethe Ansatz equations. Of course, we must leave untouched the terms really depending on the two different variables and think of them as {\it driving} or {\it forcing} terms connecting the two Hubbard models. For this reason, we move them on the l.h.s. of the equations and write
\ba
\label{firstH}
\epsilon_y^{\a}(q)+\sum_{Q=1}^{\infty}(\phi_{xy}^{Q}*L_Q)(q)&=& \sum_{M=1}^{\infty} \int_{-\infty}^{\infty}d\lambda~a_M(\lambda-\sin(q))\ln(1+e^{-\epsilon^{\a}_{v,M}(\lambda)})\nonumber\\
&-&\sum_{M=1}^{\infty}\int_{-\infty}^{\infty}d\lambda~a_M(\lambda-\sin(q))
\ln(1+e^{-\epsilon_{w,M}(\lambda)})~,~~~~~~~\label{hub1}\\
\epsilon^{\a}_{v,K}(\lambda)+\sum_{Q=1}^{\infty}(\phi_{xv}^{QK}*L_Q)(\lambda)&=&
-\int_{-\pi}^{\pi}dq~\cos(q)\,a_K(\sin(q)-\lambda)\ln(1+e^{-\epsilon_y^{\a}(q)})\nonumber\\
&+&\sum_{M=1}^{\infty}(A_{MK}*L^{\a}_{v,M})(\lambda)~,\label{hub2}\\
\epsilon^{\a}_{w,K}(\lambda)&=&-\int_{-\pi}^{\pi}dq~\cos(q)\,a_K(\sin(q)-\lambda)\ln(1+e^{-\epsilon_y^{\a}(q)})\nonumber\\
&+&\sum_{M=1}^{\infty}(A_{MK}*L^{\a}_{w,M})(\lambda)~,\label{hub3}
\ea
where

\ba
a_K(x)&=&\frac{1}{2\pi}\frac{K/g}{(K/2g)^2+x^2}\ ,\\
(A_{MK}*L)(x)&=&\int_{-\infty}^{\infty}\frac{dy}{2\pi}\,\frac{d}{dx}\,\Theta_{MK}\left(2g(x-y)\right)L(y)\ , \label{aa}\\
\Theta_{MK}(x)&=&\left\lbrace
\begin{array}{ll}
\theta(\frac{x}{|K-M|})+2\,\theta(\frac{x}{|K-M|+2})+...+2\,\theta(\frac{x}{K+M-2})+\theta(\frac{x}{K+M})\,, \mbox{if}\ K\neq M\\\\
2\,\theta(\frac{x}{2})+2\,\theta(\frac{x}{4})+...+2\,\theta(\frac{x}{2\,M-2})+\theta(\frac{x}{2\,M})\,, \mbox{if}\ K=M\ ,
\end{array}
\right.\\
\theta(x)&=&2\arctan(x)\ .
\label{lastH}
\ea
Equations (\ref{firstH}-\ref{lastH}) should be compared with equations (5.43) and (5.54-5.56) in
\cite{onedHubbard} evaluated at $\bar{u} \equiv u^{\text{Ref.} \cite{onedHubbard}}=1/2g$.

In the following sections we shall derive  a  set of functional identities (Y-system) satisfied by the quantities $Y_A=e^{\ep_A}$ (or $=e^{-\ep_A}$). Very importantly, a Y-system is universal in the sense that it  is the  same for {\it all} the  energy states $E_n(L)$, at least in a relativistic theory  ~\cite{Bazhanov:1996aq,Dorey:1996re}. Fugacities as those defined in (\ref{fugacities}) may be removed by a simple redefinition of the $Y$s. Therefore these are discharged in the next sections.

%
%
\resection{Y-system for the Hubbard model}
\label{hubbard}

The TBA equations for the Hubbard model in universal form  are
written, for example, in ~\cite{onedHubbard}~\footnote{ See also \cite{Juttner:1997tc} for the Y-system and the excited states in a closely-related model.}.  This section is not meant to be particularly
original and its aim is to explain how a subset of the Y-system equations proposed
in \cite{Gromov:2009tv} and in this paper emerges from the  Hubbard model. The TBA equations are:
\bea
\ln \eta_1(\lambda)&=& s*\ln(1+\eta_2)(\lambda) -\int_{-\pi}^{\pi} dk \cos(k) s(\lambda-\sin(k)) \ln(1+\frac{1}{\zeta(k)})~, \nn \\
\ln \eta'_1(\lambda)&=& s*\ln(1+\eta'_2)(\lambda) -\int_{-\pi}^{\pi} dk \cos(k) s(\lambda-\sin(k)) \ln(1+\zeta(k))~,\nn  \\
\ln \eta_n(\lambda)&=&s*\ln[(1+\eta_{n-1})(1+\eta_{n+1})](\lambda)~~,n=2,3,\dots~,~\nn \\
\ln \eta'_n(\lambda)&=&s*\ln[(1+\eta'_{n-1})(1+\eta'_{n+1})](\lambda)~~,n=2,3,\dots~,
\eea
and
\bea
\ln \zeta(k) &=& - {2 \over T} \cos(k) -{1 \over T} \int_{-\infty}^{\infty}
d\lambda \;s(\sin(k)-\lambda) \left( 4 \text{Re} \sqrt{1-(\lambda-i \bar{u})^2} \right) \nn \\
&+&\int_{-\infty}^{\infty} dy
\; s(\sin(k)-\lambda) \ln \left( {1 + \eta'_1 \over  1 + \eta_1} \right)~,
\eea
where
\eq
s(\lambda)= {1 \over 4 \bar{u} \cosh(\pi \lambda/2 \bar{u})}~,
\en
is the convolution kernel. $s(\lambda)$   fulfills the following important property
\eq
s(\lambda+i \bar{u})+s(\lambda-i \bar{u})= \delta(\lambda)~.
\label{prop}
\en
Relation (\ref{prop}) leads to the following  set of functional relations
\bea
\eta_n(\lambda+i \bar{u})\eta_n(\lambda-i \bar{u})=(1+\eta_{n-1}(\lambda))(1+\eta_{n+1}(\lambda))~, \\
\eta'_n(\lambda+i \bar{u})\eta'_n(\lambda-i \bar{u})=(1+\eta'_{n-1}(\lambda))(1+\eta'_{n+1}(\lambda))~,
\eea
with $n=2,3,\dots$. For $n=1$ we have instead
\bea
\ln[\eta_1(\lambda+i \bar{u})\eta_1(\lambda-i \bar{u})] &=& \ln[(1+\eta_2)(\lambda)]-\int_{-\pi}^{\pi} dk \cos(k) \delta(\lambda-\sin(k))
\ln\left(1+\frac{1}{\zeta(k)}\right)~, \nn \\
\ln[\eta'_1(\lambda+i \bar{u})\eta'_1(\lambda-i \bar{u})] &=& \ln[(1+\eta'_2)(\lambda)]-\int_{-\pi}^{\pi} dk\cos(k) \delta(\lambda-\sin(k))
\ln(1+\zeta(k))~.  \nn
\eea

But  for fixed $0 <\lambda <1$ the argument of the Dirac $\delta$
function vanishes two times, i.e.  at
$k=\arcsin(\lambda)$ and $k=\pi-\arcsin(\lambda)$. This gives
\bea
\eta_1(\lambda+i \bar{u})\eta_1(\lambda-i \bar{u})=(1+\eta_2(\lambda))
\left( { 1+1/\zeta(\pi-k) \over 1+1/\zeta(k)} \right)~,\\
\eta'_1(\lambda+i \bar{u})\eta'_1(\lambda-i \bar{u})=(1+\eta'_2(\lambda))
\left( { 1+\zeta(\pi-k) \over 1+\zeta(k)} \right)~.
\eea
Finally considering that $\cos(k)= -\sqrt{ 1- \sin^2(k)}$ for $\pi/2<k<\pi$ we get
\eq
\zeta^+(\pi-k) \zeta^-(\pi-k) \equiv \zeta(\pi-\arcsin(\lambda+i \bar{u}))\zeta(\pi-\arcsin(\lambda-i \bar{u})) =
\left({ 1+\eta'_1(\lambda) \over 1+\eta_1(\lambda)} \right)~.
\en
>From the relation
\eq
\zeta(\pi-k)=\zeta(k)e^{4 \cos(k)/T}
\label{const}
\en
(see eq. (5.A.2) in \cite{onedHubbard})
we also have
\bea
\zeta^+(k) \zeta^-(k) &\equiv&
\zeta(\arcsin(\lambda+i \bar{u}))\zeta(\arcsin(\lambda-i \bar{u})) = \left({ 1+\eta'_1(\lambda) \over 1+\eta_1(\lambda)} \right) \nn \\
&\times&  e^{\fract{4}{T} \left(\sqrt{1-(\sin(k)+i \bar{u})^2}+\sqrt{1-(\sin(k)-i \bar{u})^2} \right)}~.
\label{ex}
\eea
To see  the relationship  with the Y-system represented in figure 1 of  \cite{Gromov:2009tv}, set $z_i=1/\eta'_i$:
\bea
z_1(\lambda+i \bar{u})z_1(\lambda-i \bar{u})&=&(1+1/z_2(\lambda))^{-1}
\left({ 1+\zeta(k) \over 1+\zeta(\pi-k)} \right)~, \\
z_n(\lambda+i \bar{u})z_n(\lambda-i \bar{u}) &=&(1+1/z_{n-1}(\lambda))^{-1}(1+1/z_{n+1}(\lambda))^{-1}~,
\eea
and
\bea
Y_{2 2}(k)=\zeta(k) &,& Y_{1 1}(k) \equiv 1/Y_{22}(\pi-k)=1/\zeta(\pi-k)~, \label{rel}  \\
Y_{1,b+1}(\lambda)=z_{b}(\lambda) &,& Y_{a+1, 1}(\lambda)=\eta_{a}(\lambda)~,
\eea
with ($a,b=1,2,3,\dots$) and construct a  TBA diagram using the following rules~\cite{Quattrini:1993sm}:
\begin{itemize}
\item{ starting from a given node $(a,b)$  the l.h.s of the Y-system is
always  $Y_{ab}(\lambda+i\bar{u})Y_{ab}(\lambda-i \bar{u})$;}
\item{an horizontal link between the node $(a,b)$ and $(a',b)$ corresponds to a factor $(1+Y_{a'b}(\lambda))$ on the r.h.s. ;}
\item{a vertical link between $(a,b)$ and $(a,b')$ corresponds  to a factor $(1+1/Y_{ab'}(\lambda))^{-1}$ on the r.h.s. .}
\end{itemize}

\begin{figure} [h]
\centering
\includegraphics[width=0.35\textwidth]{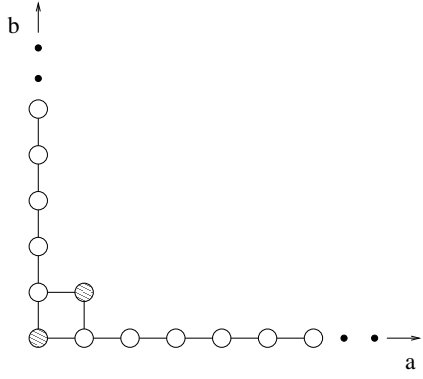}
\caption{The Hubbard diagram}
\label{fig1}
\end{figure}

It is easy to check that the diagram represented in figure~\ref{fig1} is reproduced with the
exception of the functional relation ~(\ref{ex}) for $Y_{2 2}(\lambda(k))=\zeta(k)$ which would close a `standard'  Y-system diagram only if this extra constraint were true
\eq
{ \eta_1(\lambda(k)) \over \eta_1'(\lambda(k))}= e^{\fract{4}{T} \left(\sqrt{1-(\sin(k)+i \bar{u})^2}+\sqrt{1-(\sin(k)-i \bar{u})^2} \right)}~.
\label{extra}
\en
This equation certainly holds at $T=\infty$ and would be compatible with some of  the evident symmetries of the TBA equations but still it would imply a chain of extra constraints (on the other TBA functions) that we did not try to prove. In fact, we should stress that we have included the node $Y_{11}$, which is  related to $Y_{22}$ by (\ref{rel}) and (\ref{const}). Therefore, there is no need to show an extra equation for $Y_{22}(\lambda(k))=\zeta(k)$\footnote{{\it A fortiori}, if this equation should not respect the `standard' form of the $Y$-system.}, once we already have $\ln Y_{11}(\lambda(k))=-\ln \zeta(\pi-k)$ in the TBA system.

\begin{figure} [htbp]
\centering
\includegraphics[width=0.4\textwidth]{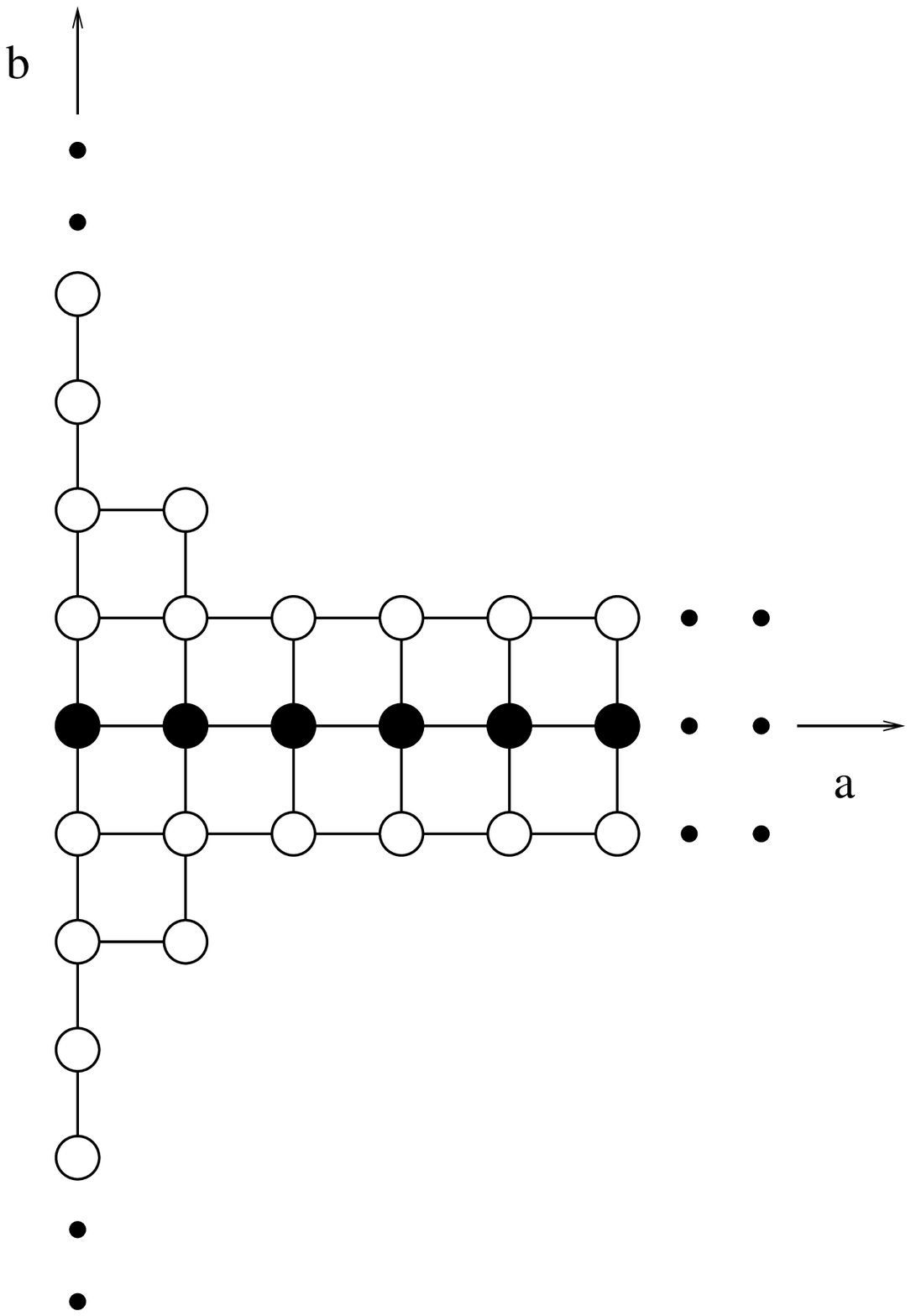}
\caption{The $AdS/CFT$ diagram}
\label{fig2}
\end{figure}

\resection{Y-system for the AdS/CFT correspondence}
\label{N4}
Let us start from equation (\ref{TBA4}) and observe that the $S_{KM}(u)$ defined in (\ref{SKM}) are a particular $n \rightarrow \infty$ limit of the $Z_n$-related  scattering matrix elements  proposed in~\cite{Koberle:1979sg}. They satisfy the following  set of functional relations~\cite{Zamolodchikov:1991et,Ravanini:1992fi}
\eq
S_{KM}\left (2\lambda+\fract{i}{g} \right)S_{KM}\left(2\lambda-\fract{i}{g} \right)= 
\prod_{K'=1}^{\infty} \left( S_{K'M}(2\lambda) \right)^{I_{K K'}}
 e^{-i2 \pi I_{K M}\Theta(2\lambda)}~,
 \label{sid}
\en
where $I_{N M}= \delta_{N,M+1}+\delta_{N,M-1} $ and $\Theta(u)$ is the Heaviside step function. Equation (\ref{sid}) leads to
\eq
\phi^{KM}_{ww}\left(\lambda'-\lambda+\fract{i}{2g} \right)+\phi^{KM}_{ww}\left(\lambda'-\lambda-
\fract{i}{2g} \right)-
\sum_{K'=1}^{\infty}  I_{K K'} \phi^{K'M}_{ww}(\lambda'-\lambda)= -I_{K M}\delta(\lambda'-\lambda)~.
\label{k1prop}
\en
Notice that $\phi^{KM}_{ww}(\lambda)$ is equal to   $-A_{KM}(\lambda)$ defined in  equation (\ref{aa}). Another relevant identity is
\bea
\phi^K_{yw}\left(\sin(q'),\lambda+\fract{i}{2g} \right)&+&
\phi^K_{yw}\left(\sin(q'),\lambda-\fract{i}{2g} \right)- \nn \\
\sum_{K'=1}^{\infty}  I_{K K'} \phi^{K'}_{yw}(\sin(q'),\lambda) &=& -\delta_{K1}\cos(q') \delta(\sin(q')-\lambda)~.
\label{k2prop}
\eea
Using equations  (\ref{k1prop}), (\ref{k2prop}) and setting
\eq
Y_{w,K}^{\alpha}(\lambda)=e^{-\ep_{w,K}^{\alpha}(\lambda)}~~,~Y_{y}^{\alpha}(q)=e^{\ep_{y}^{\alpha}(q)}~~,~
Y_{y^*}^{\alpha}(q) \equiv e^{\ep_{y^*}^{\alpha}(q)}= e^{-\ep_{y}^{\alpha}(\pi-q)}~,
\en
with $q=\arcsin(\lambda)$, we find
\eq
Y_{w,K}^\alpha(\lambda+\fract{i}{2g})Y_{w,K}^\alpha(\lambda-\fract{i}{2g}) = \prod_{K'=1}^{\infty}\left(1+{1 \over Y_{w,K'}^\alpha(\lambda)} \right)^{-I_{KK'}} \left({ 1 + Y_{y^*}^\alpha(q) \over 1 + 1/Y^{\alpha}_y(q) } \right)^{\delta_{K1}}~.
\label{Ys1}
\en
Let us now consider equation (\ref{TBA3}). The  identity (\ref{eqrt})
with $K=2,3,\dots$, together with equations (\ref{k1prop}) and (\ref{k2prop}) lead to

\eq
Y^{\alpha}_{v,K}(\lambda+\fract{i}{2g})Y^{\alpha}_{v,K}(\lambda-\fract{i}{2g}) = \prod_{K'=1}^{\infty}(1+Y^{\alpha}_{v,K'}(\lambda))^{I_{K K'}}
\left( 1 + {1 \over Y_{K+1}(\tilde{p})} \right)^{-1}
\label{Ys2}~,
\en
with $\tilde{p}=\tilde{p}(2 \lambda)$ defined  in (\ref{pt}) and $Y^{\alpha}_{v,K}=e^{\ep_{v,K}^{\alpha}}$.
The case with $K=1$ is slightly more tricky, but the game is just  the same.
One starts considering the expression
\eq
\ep_{v1}^{\alpha}(\lambda+\fract{i}{2g})+\ep_{v1}^{\alpha}(\lambda-\fract{i}{2g})
-\ep_{v2}^{\alpha}(\lambda)- \ep_{y}^{\alpha}(q)-\ep_{y^*}^{\alpha}(q)~,
\en
with  $q=\arcsin(\lambda)$. The corresponding r.h.s. of the TBA equations cancel almost completely due to the functional relations
fulfilled by the kernel functions, they just leave some `contact' delta function
contributions. In this case the result is
\bea
Y^{\alpha}_{v,1}(\lambda+\frac{i}{2g})Y^{\alpha}_{v,1}(\lambda-\frac{i}{2g})&=&
(1+Y^{\alpha}_{v,2}(\lambda))
(1+Y^{\alpha}_{y}(q)) \nn \\
&\times&
\left( 1 + {1 \over Y_{y^*}(q)} \right)^{-1}
\left( 1 + {1 \over Y_2(\tilde{p})} \right)^{-1}
\label{Ys3}~.
\eea
Further, consider the quantity
\eq
\ep_{y}(q^+)+\ep_{y}(q^-)-\ep_{v,1}(\lambda)~,
\en
where $q^{\pm}= \arcsin(\lambda\pm i/2g)$,  the kernel  properties and  the TBA equation (\ref{TBA4})
 at $K=1$ give
\eq
Y^{\alpha}_y(q^+) Y^{\alpha}_y(q^-) =(1+Y^{\alpha}_{v,1}(\lambda))\left( 1 + {1 \over Y^{\alpha}_{w,1}(\lambda)} \right)^{-1}
\left( 1 + {1 \over Y_1(\tilde{p})} \right)^{-1}~,
\label{Ys4}
\en
with $\tilde{p}=\tilde{p}(2 \lambda)$. Finally, using  the property
\eq
{x^{Q+}(u+i/g) \over x^{Q-}(u+i/g)}{x^{Q+}(u-i/g) \over x^{Q-}(u-i/g)}
=
{x^{(Q-1)+}(u) \over x^{(Q-1)-}(u)}{x^{(Q+1)+}(u) \over x^{(Q+1)-}(u)}~,
\en
and similar relations for $\phi^{Q'Q}_{\sl(2)}$, $\phi_{yx}^Q$ and $\phi_{vx}^{QM}$ (see Appendix~B) we get
\eq
Y_{Q}(x(u+\fract{i}{g}))Y_{Q}(x(u-\fract{i}{g})) = \prod_{Q'=1}^{\infty}(1+Y_{Q'}(x(u)))^{I_{Q Q' }}
\prod_{\alpha=1}^2 \left( 1 + {1 \over Y_{v,Q-1}^{\alpha}(\lambda)} \right)^{-1}~,
\label{Ys5}
\en
with $Q=2,3,\dots$ and
\eq
Y_{1}(x(u+\fract{i}{g}))Y_{1}(x(u-\fract{i}{g})) = (1+Y_{2}(x(u)))
\prod_{\alpha=1}^2 \left( 1 + {1 \over Y^\alpha_y(q)} \right)^{-1}~,
\label{Ys6}
\en
with $q= \arcsin(\lambda)$,  $u=2 \lambda$ and $Y_Q=e^{\ep_Q}$.
Setting
\bea
Y_{Q,0}=Y_Q&,&Y_{1,1}=Y_y^{1}~~,~Y_{1,-1}=Y_y^{2}~~,~Y_{2,2}=Y_{y^*}^{1}~~,~Y_{2,-2}=Y_{y^*}^{2}~~,~~~\nn \\
Y_{1,K+1}=Y^{1}_{w,K}&,&Y_{1,-K-1}=Y^{2}_{w,K}~~,~
Y_{K+1,1}=Y^{1}_{v,K}~~,Y_{K+1,-1}=Y^{2}_{v,K}~,
\label{id}
\eea
and following the rules given at the end of section~\ref{hubbard} we may encode
this Y-system  in the  diagram in figure~\ref{fig2}. In other words,  the equations  (\ref{Ys1}-\ref{Ys6})  with the identifications~(\ref{id}) can be recast in the compact form
\eq
Y_{a,b}^{+}Y_{a,b}^{-}= (1+Y_{a+1,b})(1+Y_{a-1,b})\left(1+{1 \over Y_{a,b+1}} \right)^{-1}
\left(1+{1 \over Y_{a,b-1}} \right)^{-1}~,~~
\en
as long as $(a,b) \ne (2,\pm 2)$.

Our $Y$-diagram shares its structure with that in figure 1 of~\cite{Gromov:2009tv}. Yet, we shall remark the exact parallel to what we have noticed at the end of section~\ref{hubbard} about the Hubbard model: to close completely the  diagram by using the `standard' rules, we would need two extra equations
\eq
Y^{\alpha}_{y^*}(q^{+}) Y^{\alpha}_{y^*}(q^{-}) =\left( 1 +  Y^{\alpha}_{w,1}(\lambda) \right)\left(1+{1 \over Y^{\alpha}_{v,1}(\lambda)} \right)^{-1}~,~~(\alpha=1,2)~.
\label{extra2}
\en
The careful reader may have noticed that we did not  prove these equations, since for  the nodes $(2,\pm 2)$ we already have the identification
\eq
Y^{\alpha}_{2,\pm 2}(q)= {1 \over Y^{\alpha}_{1,\pm 1}(\pi-q)}~~~,
\en
and thus, at any rate, we do not need to include the associated equations  in the TBA system.
A careful analysis suggests that equation (\ref{extra2}) is in general incorrect.

\section{Partial conclusions and remarks}

In a nutshell, we have proposed the TBA equations which should control the energy/dimension spectrum of the $\text{AdS}_5\times\text{S}^5$ correspondence. We have also derived from them the universal $Y$-system which should characterise any state of the theory for any value of the coupling constant $g$. Of course, since universal, this system contains the information about a specific state in a much more involved way.

Nevertheless, we may still lean on the theory of massive integrable field theories. In this area a clear procedure has been established to extract  excited state  non-linear integral equations from that of  the ground state: this is initially described from three different perspectives in the  papers \cite{Bazhanov:1996aq, Dorey:1996re, FMQR} . Essentially, it proves the recipe to extract suitable driving terms $\sum_i \ln S(u_i,u)$ as residues of the convolution integrals, and these terms clearly involve the scattering matrix elements. Under the perspective of the non-linear integral equation, this idea has been already applied to some sectors of the asymptotic Beisert-Staudacher equations \cite{FR, FRS, BFR}.

Re-echoing the title of \cite{FFGR}, the Hubbard model excursion seems to be still on in this discipline. In fact, we have found just two copies of this model, talking through their massive nodes. Moreover, this is also the structure of the $Y$-system recently proposed by Gromov, Kazakov and Viera (somehow on symmetry grounds) \cite{Gromov:2009tv}.

Despite the lack of a BA or integrability description of sufficiently {\it short} operators, we may consider all these arguments in favour of a TBA description of the correspondence.

\section*{Acknowledgements}
DF is particularly indebted to M. Rossi for insightful discussions and suggestions. We also thank G. Arutyunov and F. Ravanini. We acknowledge the INFN grants IS PI14{\it ``Topics in non-perturbative gauge dynamics in field and string theory''} and PI11 for travel financial support, and the University PRIN 2007JHLPEZ ``Fisica Statistica dei Sistemi Fortemente Correlati all'Equilibrio e
Fuori Equilibrio: Risultati Esatti e Metodi di Teoria dei Campi".


\resection{Appendix~A}

Here we report the definitions used for the kernels involved in the TBA equations (\ref{TBA1})-(\ref{TBA4}):

\ba
\phi_{\sl(2)}^{Q'Q}(\tilde{p}',\tilde{p})&=&\frac{1}{2\pi i}\frac{d}{d\tilde{p}'}\ln S_{\sl(2)}^{Q'Q}(\tilde{p}',\tilde{p})~,\\
\phi_{xy}^{Q}(\tilde{p},q)&=&\frac{1}{2\pi i}\frac{d}{d\tilde{p}}\ln\left(\frac{x^{Q-}(\tilde{p})-y(q)}{x^{Q+}(\tilde{p})-y(q)}\sqrt{\frac{x^{Q+}(\tilde{p})}{x^{Q-}(\tilde{p})}}\right)~,\\
\phi_{xv}^{QM}(\tilde{p},\lambda)&=&\frac{1}{2\pi i}\frac{d}{d\tilde{p}}\ln
S_{xv}^{QM}(\tilde{p},\lambda)~,\\
\phi_{vx}^{KQ}(\lambda,\tilde{p})&=&-\frac{1}{2\pi i}\frac{d}{d\lambda}\ln S_{xv}^{QK}(\tilde{p},\lambda)~,\\
\phi_{yx}^{Q}(q,\tilde{p})&=&\frac{1}{2\pi i}\frac{d}{dq}\ln \left(\frac{y(q)-x^{Q-}(\tilde{p})}{y(q)-x^{Q+}(\tilde{p})}\sqrt{\frac{x^{Q+}(\tilde{p})}{x^{Q-}(\tilde{p})}}\right)~,\\
\phi_{yv}^{K}(q,\lambda)=\phi_{yw}^{K}(q,\lambda)&=&\frac{1}{2\pi i}\frac{d}{dq}\ln \left(\frac{v(q)-2\lambda-iK/g}{v(q)-2\lambda+iK/g}\right)~,\\
\phi_{vv}^{MK}(\lambda',\lambda)=\phi_{ww}^{MK}(\lambda',\lambda)&=&\frac{1}{2\pi i}\frac{d}{d\lambda'}\ln S_{MK}(2\lambda'-2\lambda)~, \\
\phi_{vy}^{K}(\lambda,q)=-\phi_{wy}^{K}(\lambda,q)&=&\frac{1}{2\pi i}\frac{d}{d\lambda}\ln\left(\frac{2\lambda-v(q)+iK/g}{2\lambda-v(q)-iK/g}\right)~,
\ea
where
\ba
&&x^{Q\pm}(\widetilde p)= {1\ov
2g}\left(\sqrt{1 +{4g^2\ov Q^2+\widetilde p^2}}\ \mp\ 1\right)\left(
\widetilde p -i Q\right)\, ,\\
&&\tilde{p}(u)= \frac{i g}{2} \left( \sqrt{4 - \left(u+ i \fract{Q}{g}\right)^2}
-\sqrt{4 - \left(u- i \fract{Q}{g}\right)^2} \right)~, \label{pt}\\
&&y(q)=i\,e^{-iq}\ ,\ \ \ \ \ \ \ \ v(q)=2\sin(q)\ ,~~~w(\lambda)=2 \lambda~,\\
&&v_K^{\pm}(\lambda)=2\lambda_{v,K}\pm\frac{iK}{g}\ ,\ \ \ \ \ \ w_K^{\pm}(\lambda)=2\lambda_{w,K}\pm\frac{iK}{g}\ , \\
&& x(u)={1 \over 2} \left( u - i \sqrt{4 -u^2} \right)~,x^{Q\pm}(-u)= - {1 \over x^{Q\mp}(u)}~~,\\
&& x^{Q\pm}(u)=x(u \pm i \fract{Q}{g})~,~~~\tilde{p}(-u)=-\tilde{p}(u)~.
\ea
It is easy to notice that some of these kernels depends only on the difference of the rapidities, as in the relativistic case. They are

\ba
\phi_{vy}^{M}(\lambda,q)&=&-\phi_{wy}^{M}(\lambda,q)=\phi_M(\lambda-\sin(q))~,\ \mbox{where}\ \ \phi_M(\lambda)=\frac{1}{2\pi i}\frac{d}{d\lambda}\ln\left(\frac{\lambda+iM/2g}{\lambda-iM/2g}\right)~~~~~~~\\
\phi_{vv}^{MK}(\lambda',\lambda)&=&\phi_{ww}^{MK}(\lambda',\lambda)=\phi_{MK}(\lambda'-\lambda)~,\ \ \mbox{where}\ \ \phi_{MK}(\lambda)=\frac{1}{2\pi i}\frac{d}{d\lambda}\ln S_{MK}(2\lambda)~.
\ea
%
\vspace{1cm}

%

\resection{Appendix~B}
\label{Appendix2}

Here we want to show how also the other kernels satisfy an identity of the type (\ref{k1prop}). As long as the  kernel

\ba
&&\phi_{\sl(2)}^{QQ'}(u,u')=\frac{1}{2\pi i}\frac{d}{d\tilde{p}}\ln\left[\left( {u-u' + i\frac{|Q-Q'|}{g} \over u-u' -i\frac{|Q-Q'|}{g}} \right)\left( {u-u' + i\frac{Q+Q'}{g} \over u-u' -i \frac{Q+Q'}{g}} \right)\right.\nn\\
&&\left.\left[\frac{1-\frac{1}{x_k^{Q+}x_l^{Q'-}}}{1-\frac{1}{x_k^{Q-}x_l^{Q'+}}}\sigma(x^{Q\pm}_k,x^{Q'\pm}_l)\right]^{-2}\prod_{k=1}^{\text{min}(Q,Q')-1} \left( {u-u' + i\frac{|Q-Q'|+2k}{g} \over u-u' -i\frac{|Q-Q'|+2k}{g}}\right)^2\right]
\ea
is concerned, we may shift on the second variable
\ba
&&\phi^{QQ'}_{\sl(2)}\left(u,u' +\fract{i}{g}\right)+\phi^{QQ'}_{\sl(2)}\left(u,u'-\fract{i}{g}\right)=\frac{1}{2\pi i}\frac{d}{d\tilde{p}}\Bigg[\ln\left( {u -u'+ i\frac{|Q-Q'|}{g}-\frac{i}{g}} \over u-u' -i\frac{|Q-Q'|}{g}-\frac{i}{g} \right)\nn\\
&&+\ln \left( {u -u'+ i\frac{Q+Q'-1}{g}} \over u -u'-i \frac{Q+Q'+1}{g} \right)+2\sum_{k=1}^{\text{min}(Q,Q'-1)-1}\ln \left( {u -u'+ i\frac{|Q-Q'|+2k}{g}-\frac{i}{g}} \over u -u'-i\frac{|Q-Q'|+2k}{g}-\frac{i}{g}\right)\nn\\
&&+\ln\left( {u -u'+ i\frac{|Q-Q'|}{g}+\frac{i}{g}} \over u-u' -i\frac{|Q-Q'|}{g}+\frac{i}{g} \right)+\ln
\left( {u -u'+ i\frac{Q+Q'+1}{g} \over u -u'-i \frac{Q+Q'-1}{g}} \right)\nn\\
&&+2\sum_{k=1}^{\text{min}(Q,Q'+1)-1}\ln \left( {u -u'+ i\frac{|Q-Q'|+2k}{g}+\frac{i}{g}} \over u-u' -i\frac{|Q-Q'|+2k}{g}+\frac{i}{g} \right)-2\ln\left(\frac{1-\frac{1}{x\left(u-\frac{iQ}{g}\right)x\left(u+i\frac{Q'+1}{g}\right)}}{1-\frac{1}{x\left(u+\frac{iQ}{g}\right)x\left(u-i\frac{Q'-1}{g}\right)}}\right)\nn\\
&&-2\ln\left(\frac{1-\frac{1}{x\left(u-\frac{iQ}{g}\right)x\left(u+i\frac{Q'-1}{g}\right)}}{1-\frac{1}{x\left(u+\frac{iQ}{g}\right)x\left(u-i\frac{Q'+1}{g}\right)}}\right)\nn\\
&&-2i\sum _{r=2}^{\infty}\sum _{\nu =0}^{\infty} \beta
_{r,r+1+2\nu}(g)
[q_r^{Q}(u)q_{r+1+2\nu}^{Q'+1}(u')-q_r^{Q'+1}(u')q_{r+1+2\nu}^{Q}(u)\nonumber\\
&&+q_r^{Q}(u)q_{r+1+2\nu}^{Q'-1}(u')-q_r^{Q'-1}(u')q_{r+1+2\nu}^{Q}(u)]\Bigg]   ~,
\ea
so proving the identity used in the main text
\eq
\phi^{QQ'}_{\sl(2)}\left(u,u'+\fract{i}{g} \right)+\phi^{QQ'}_{\sl(2)}\left(u,u'-\fract{i}{g} \right)=\sum_{Q''=1}^{\infty} I_{Q'Q''}\phi_{\sl(2)}^{QQ''}(u,u')
-\delta(u-u')I_{QQ'}~.
\en
The bound state charges $q_r^Q$ are defined as usual \cite{Roiban:2006gs, Chen:2006gq} and the shifted charges we use above are
\be
q_r^{Q\pm1}(u)=\frac {i}{r-1} \left [ \left (\frac {1}{x\left(u+\frac{i(Q\pm1)}{g}\right)}\right
)^{r-1}-\left (\frac {1}{x\left(u-\frac{i(Q\mp1)}{g}\right)}\right )^{r-1} \right ]~.
\ee
Analogously, by direct computation
\ba
&&\phi^{MQ}_{vx}\left(\lambda,x^{Q\pm}(u+\fract{i}{g}) \right)+\phi^{MQ}_{vx}\left(\lambda,x^{Q\pm}(u-\fract{i}{g}) \right)=-\frac{1}{2 \pi i}\frac{d}{d\lambda}\left[\ln\left(\frac{x^{(Q-1)-}-x(v+\frac{iM}{g})}{x^{(Q+1)+}-x(v+\frac{iM}{g})}\right)\right.\nn\\
&&+\ln\left(\frac{x^{(Q+1)-}-x(v-\frac{iM}{g})}{x^{(Q-1)+}-x(v-\frac{iM}{g})}\right)+\ln\left(\frac{x^{(Q-1)-}-x(v-\frac{iM}{g})}{x^{(Q+1)+}-x(v-\frac{iM}{g})}\right)+\ln\left(\frac{x^{(Q+1)+}}{x^{(Q-1)-}}\right)\nn\\
&&+\ln\left(\frac{x^{(Q-1)+}}{x^{(Q+1)-}}\right)+\ln\left(\frac{x^{(Q+1)-}-x(v+\frac{iM}{g})}{x^{(Q-1)+}-x(v+\frac{iM}{g})}\right)+\sum _{j=1}^{M-1}
\left[\ln\left(\frac{u-i\frac{Q-1}{g}-(v-\frac{iM}{g})-\frac{2i}{g}j}{u+i\frac{Q+1}{g}-(v+\frac{iM}{g})+\frac{2i}{g}j}\right)\right.\nonumber\\
&&+\left.\ln\left(\frac{u-i\frac{Q+1}{g}-(v-\frac{iM}{g})-\frac{2i}{g}j}{u+i\frac{Q-1}{g}-(v+\frac{iM}{g})+\frac{2i}{g}j}\right)\right] ~,
\ea
($v=2 \lambda$ ),we may prove an identity with the same form, but involving $\phi_{vx}^{MQ}$
\bea
\phi^{MQ}_{vx}\left(\lambda,x^{Q\pm}(u+\fract{i}{g}) \right)+\phi^{MQ}_{vx}\left(\lambda,x^{Q\pm}(u-\fract{i}{g}) \right) &=&\sum_{Q'=1}^{\infty} I_{QQ'} \phi_{vx}^{MQ'}\left(\lambda,x^{Q'}(u) \right) \nn \\
&+&\delta(\lambda-u/2) \delta_{Q-1,M}~.
\eea
An identity with the same form may be derived for $\phi_{xv}^{QM}$:
\ba
&&\phi^{QM}_{xv}\left(x^{Q\pm}(u),\lambda +\fract{i}{2g}\right)+\phi^{QM}_{xv}\left(x^{Q\pm}(u)\lambda-\fract{i}{2g}\right)=\frac{1}{2 \pi i}\frac{d}{d\tilde{p}}\left[\ln\left(\frac{x^{Q-}-x\left(v+\frac{i(M+1)}{g}\right)}{x^{Q+}-x\left(v+\frac{i(M+1)}{g}\right)}\right)\right.\nn\\
&&+\ln\left(\frac{x^{Q-}-x\left(v-\frac{i(M-1)}{g}\right)}{x^{Q+}-x\left(v-\frac{i(M-1)}{g}\right)}\right)+\ln\left(\frac{x^{Q-}-x\left(v-\frac{i(M+1)}{g}\right)}{x^{Q+}-x\left(v-\frac{i(M+1)}{g}\right)}\right)+2\ln\left(\frac{x^{Q+}}{x^{Q-}}\right)\nn\\
&&+\ln\left(\frac{x^{Q-}-x\left(v+\frac{i(M-1)}{g}\right)}{x^{Q+}-x\left(v+\frac{i(M-1)}{g}\right)}\right)+\sum _{j=1}^{M-1}
\left[\ln\left(\frac{u-i\frac{Q}{g}-\left(v-i\frac{M-1}{g}\right)-\frac{2i}{g}j}{u+i\frac{Q}{g}-\left(v+i\frac{M+1}{g}\right)+\frac{2i}{g}j}\right)\right.\nonumber\\
&&+\left.\ln\left(\frac{u-i\frac{Q}{g}-\left(v-i\frac{M+1}{g}\right)-\frac{2i}{g}j}{u+i\frac{Q}{g}-\left(v+i\frac{M-1}{g}\right)+\frac{2i}{g}j}\right)\right] ~,
\eea
\bea
\phi^{QM}_{xv}\left(x^{Q\pm}(u),\lambda +\fract{i}{2g}\right)+\phi^{QM}_{xv}\left(x^{Q\pm}(u),\lambda-\fract{i}{2g}\right)&=&\sum_{M'=1}^{\infty} I_{MM'} \phi_{xv}^{QM'}\left(x^{Q}(u),\lambda \right) \nn \\
&+&\delta(\lambda-u/2) \delta_{Q-1,M}~.
\label{eqrt}
\ea

%
%
%
%
%
\end{document}